\documentstyle[sevensterapj,sevensterpsfig]{sevensterart}

\def\shortauthors{\lefthead}
\def\shorttitle{\righthead}

\newcount\levelone    \levelone=0
\newcount\leveltwo    \leveltwo=0
\newcount\levelthree  \levelthree=0
\newcount\levelfour   \levelfour=0
\def\chaphead{}                             
\def\secno{\chaphead\the\levelone}
\def\subno{\chaphead\the\levelone.\the\leveltwo}
\def\subsubno{\chaphead\the\levelone.\the\leveltwo.\the\levelthree}
\def\subsubsubno{\chaphead\the\levelone.\the\leveltwo.\the\levelthree
                           .\the\levelfour}
\def\newsec{\advance\levelone by1 \leveltwo=0 \levelthree=0 \levelfour=0}
\def\newsub{\advance\leveltwo by1 \levelthree=0 \levelfour=0}
\def\newsubsub{\advance\levelthree by1 \levelfour=0}
\def\newsubsubsub{\advance\levelfour by1}

\def\absnarrower{\advance\leftskip by \abstractindent}

\newdimen\secskipamount  \secskipamount=1pt
\newdimen\subskipamount  \subskipamount=1pt
\newdimen\bottomtol \bottomtol=0.03\vsize
\def\secskip{\par \ifdim\lastskip<\secskipamount \removelastskip \fi
    \vskip 0pt plus \bottomtol \penalty-250
    \vskip 0pt plus -\bottomtol \relax
    \vskip\secskipamount plus3pt minus3pt}
\def\subskip{\par \ifdim\lastskip<\subskipamount \removelastskip \fi
    \vskip 0pt plus 0.5\bottomtol \penalty-150
    \vskip 0pt plus -0.5\bottomtol \relax
    \vskip\subskipamount plus2pt minus2pt}
\def\subsubskip{\par \ifdim\lastskip<\subskipamount \removelastskip \fi
    \vskip 0pt plus 0.5\bottomtol \penalty-150
    \vskip 0pt plus -0.5\bottomtol \relax
    \vskip\subskipamount plus2pt minus2pt \hskip 10pt}

\outer\def\unnumberedsectionbegin #1 #2\par {\secskip \noindent {{\bf  #1}
\dotfill #2}
    \nobreak \vskip 1pt \noindent}

\outer\def\sectionbegin #1 #2\par {\secskip \newsec \noindent {{\bf \secno\  #1}
\dotfill #2}
    \nobreak \vskip 1pt \noindent}
\outer\def\subsectionbegin #1 #2\par {\subskip \newsub {\subno\ {\rm #1} \hfill
#2}
    \nobreak \vskip 1pt \noindent}
\outer\def\subsubsectionbegin #1 #2\par {\subsubskip \newsubsub
    {\subsubno\ {\it #1} \hfill #2}
  \nobreak \vskip 1pt \noindent}

\newcount\eqnumber \eqnumber=0
\def\new{{\rm\chaphead\the\eqnumber}\global\advance\eqnumber by 1}

\newcount\fignumber \fignumber=0
\def\nfig{\chaphead\the\fignumber\global\advance\fignumber by 1}
\def\ntab{\chaphead\the\tabnumber\global\advance\tabnumber by 1}

\newcount\fononum \fononum=0
\def\nfn{\global\advance\fononum by 1}
\def\fonono{\the\fononum}

\def\bck{\hskip-0.35em}
\def\wisk#1{\ifmmode{#1}\else{$#1$}\fi}
\def\etal{{et al.$\,$}}
\def\mlod{micro--lensing optical depth}

\def\lvd{longitude--velocity diagram}

\def\losa{line--of--sight}
\def\losn{line of sight}

\def\fv{\wisk{ f_{\rm V}}}
\def\gc{galactic Centre}

\def\msol{\wisk{\,\rm M_\odot}}

\let\rsun=\rsol

\def\degr{\wisk{^{\circ}}}                                
\let\deg=\degr
\def\decdeg#1.#2 {\wisk{#1^{\,\rm o}\bck.\,#2}\ }
\def\decmin#1.#2 {\wisk{#1^{\,\prime}\bck.\,#2}\ }

\def\decsec#1.#2 {\wisk{#1^{\prime\prime}\hskip-0.42em.\hskip0.10em#2}\ }

\def\kms{\wisk{\,\rm km\,s^{-1}\,}}                    

\def\oversim#1#2{\lower1.5pt\vbox{\baselineskip0pt \lineskip-0.5pt
     \ialign{$\mathsurround0pt #1\hfil##\hfil$\crcr#2\crcr\sim\crcr}}}

\def\eqref#1{\advance\eqnumber by -#1 \chaphead\the\eqnumber
           \advance\eqnumber by #1 }
\def\?{\eqref{1}}
\def\last{\advance\eqnumber by -1 {\rm\chaphead\the\eqnumber}\advance
     \eqnumber by 1}
\def\eqnam#1{\xdef#1{\chaphead\the\eqnumber}}
\def\appendixbegin#1 #2{\eqnumber=1 \def\chaphead{{#1}}
    \levelone=0\leveltwo=0\levelthree=0\levelfour=0\eqnumber=1\fignumber=1
    \vskip\subskipamount\noindent{\ninepoint\bf Appendix #1\ \ \ #2}
    \vskip\subskipamount\noindent}
\def\noappendixbegin#1 #2{\eqnumber=1 \def\chaphead{{#1} }
    \levelone=0\leveltwo=0\levelthree=0\levelfour=0\eqnumber=1\fignumber=1
    \vskip\subskipamount\noindent{}
    \vskip\subskipamount\noindent}
\def\nfig{\chaphead\the\fignumber\global\advance\fignumber by 1}
\def\anfig{\global\advance\fignumber by 1}

\def\ntab{\chaphead\the\tabnumber\global\advance\tabnumber by 1}
\def\antab{\global\advance\tabnumber by 1}
\def\nfiga#1{\chaphead\the\fignumber{#1}\global\advance\fignumber by 1}
\def\rfig#1{\advance\fignumber by -#1 \chaphead\the\fignumber
            \advance\fignumber by #1}
\def\fignam#1{\xdef#1{\chaphead\the\fignumber}}
\def\tabnam#1{\xdef#1{\chaphead\the\tabnumber}}

\def\aa#1 #2 {, {A\&A,}{ #1, #2} }
\def\aal#1 #2 {, {A\&A,}{ #1, L#2}\ }
\def\aas#1 #2 {, {A\&AS,}{ #1, #2} }
\def\aj#1 #2 {, {AJ, }{#1, #2}\ }
\def\apj#1 #2 {, {ApJ, }{#1, #2}\ }
\def\apjl#1 #2 {, {ApJ, }{#1, L#2 }\ }
\def\apjs#1 #2 {, {ApJS, }{#1, #2}\ }
\def\araa#1 #2 {, {ARA\&A, }{#1, #2}\ }
\def\mnras#1 #2 {, {MNRAS, }{#1, #2}\ }
\def\bargal#1 { {1996, In: {Buta, R., Crocker, D., Elmegreen, B.~(eds.)
      Barred Galaxies, PASPC 91, San Francisco,} p. #1 }}
\def\thrss#1 { {1999, In: Gibson, B., Axelrod, T., Putman, M.~(eds.)
     The Third Stromlo Symposium: The Galactic Halo,
     PASPC 165, San Francisco, p. #1 }}

\newcount\fignumber \fignumber=1
\newcount\eqnumber \eqnumber=1
\newcount\tabnumber \tabnumber=1

\def\Sct{Section\ }
\def\Eqt{equation\ }
\def\Fg{Fig.\ }

\def\DTAU{1}
\def\DMAS{2}
\def\DTDM{3}

\def\PARS{1}

\def\GALS{1}
\def\HIYY{2}
\def\BWF{3}
\def\FTE{4}
\def\CFTE{5}
\def\STRM{6}
\def\DMLR{7}

\begin{document}

\def\bibitm{\bibitem{}}

\title{An inner ring and the micro lensing toward the Bulge}

\author{Maartje N.~Sevenster\altaffilmark{1}
\ and Agris J.~Kalnajs\altaffilmark{2}}

\altaffiltext{1}{msevenst@mso.anu.edu.au}
\altaffiltext{2}{agris@mso.anu.edu.au}
\affil{MSSSO/RSAA, Cotter Road, Weston ACT 2611, Australia}

\shortauthors{M.~Sevenster and A.~Kalnajs}
\shorttitle{Inner ring and micro lensing}

\begin{abstract}
All current Bulge--Disk models for the inner Galaxy fall short of
reproducing self--consistently 
the observed \mlod\ by a factor of two ($> 2\sigma$).
We show that the least mass--consuming way to increase 
the \mlod\ is to add density roughly half--way the 
observer and the highest micro--lensing--source density.
We present evidence for the existence of such a density
structure in the Galaxy: an inner
ring, a standard feature of barred galaxies. 
Judging from data on similar rings in external galaxies,
an inner ring can contribute more than 50\% 
of a pure Bulge--Disk model to the \mlod .
We may thus eliminate the need for a small viewing angle of
the Bar. The influence of an inner ring
on the event--duration distribution, for realistic viewing
angles, would be to increase the fraction of long--duration events
toward Baade's window. The longest events are expected
toward the negative--longitude tangent point at $\ell$$\sim$$-$22\degr .
A properly sampled event--duration distribution
toward this tangent point would provide 
essential information about viewing angle and elongation
of the over--all density distribution in the inner Galaxy.
\end{abstract}



\section{Introduction}

The ``micro--lensing optical depth'' as a function of position on the sky 
is the probability that a given star in that direction 
will be gravitationally lensed by a foreground object 
at a given instant.
The \mlod\ $\tau$ measured toward the Bulge (or Baade's window)
(the MACHO project, Alcock \etal1997; the OGLE project,
Udalski \etal1994; the DUO project, Alard \& Guilbert 1997)
provides important information about the stellar and
dark--matter distribution in the inner Galaxy. 

Other than for the Magellanic Clouds, one 
usually tries to explain the Bulge's
\mlod\ with ``self--lensing'' models; 
the lenses and the sources come from the same (dynamical) distribution.
Justification for this comes from the
micro--lensing event--duration distribution, as well
as recent determinations of the initial--mass function and
the fraction of invisible matter in the solar neighbourhood,
all of which indicate that there is no significant amount
of dark matter in the plane of the Galaxy nor a large number
of brown dwarfs
(Alcock \etal1997; Gould, Flynn \& Bahcall 1998; Kroupa 1998; 
Kuijken \& Gilmore 1989; Fuchs, Jahreiss \& Flynn 1998).
Note that even for the lines of sight toward the LMC,
in fact, dark matter plays a minor 
role in microlensing (Alcock \etal2000b).

\begin{figure*}
\fignam\GALS
\anfig
{\psfig{figure=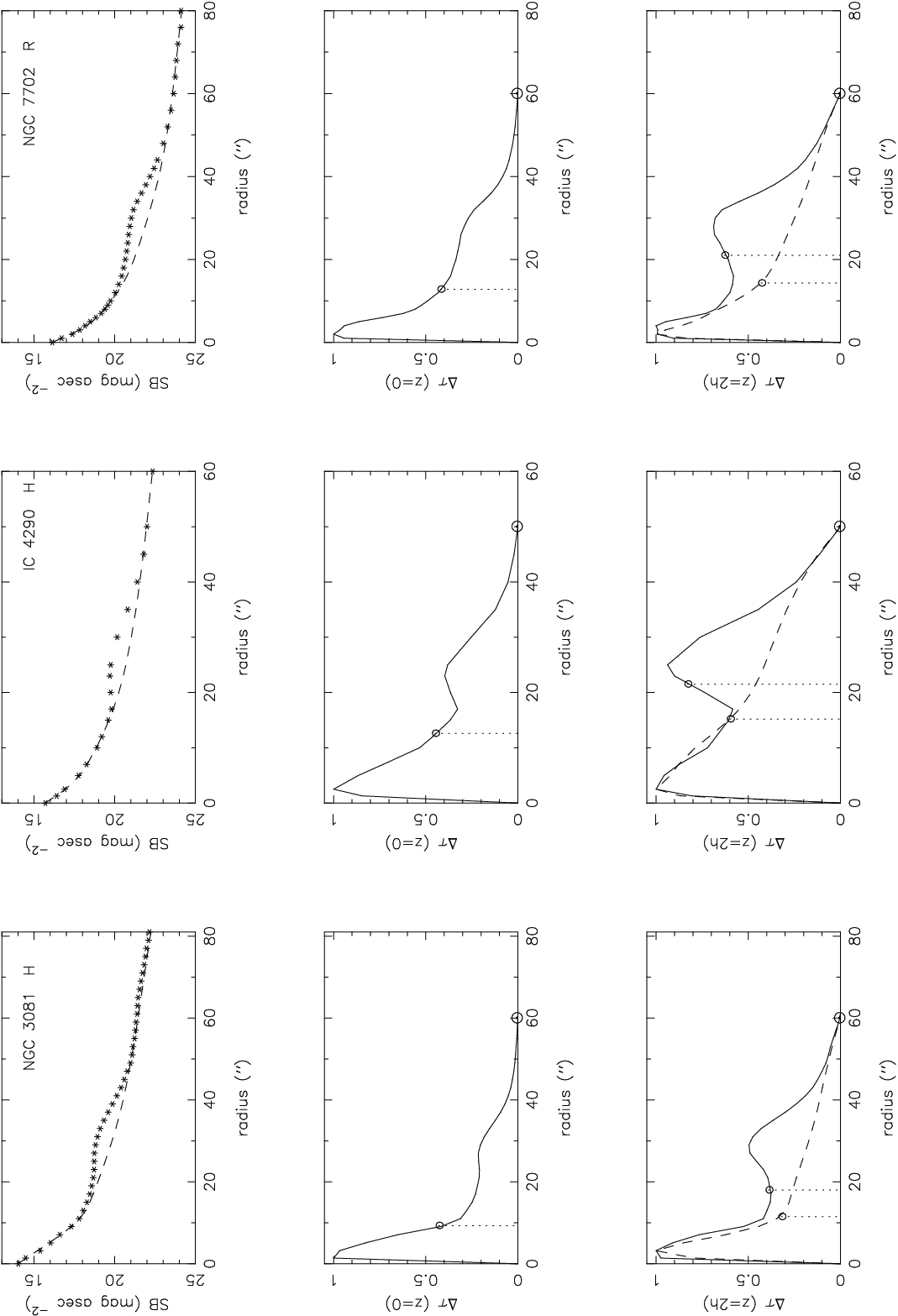,width=17truecm,angle=270}}
\figcaption{
Elliptically averaged surface--density
profile (dots) for the galaxy NGC 3081 (Buta \& Purcell 1998), 
IC 4290 (Buta \etal1998) and NGC 7702 (Buta 1991)
in magnitudes per square arcsecond (top panels).
The middle panels show the contribution, $\Delta \tau$, 
to the \mlod\ along the \losn\ towards a source in the
centre of the galaxies.
The bottom panels give $\Delta \tau$ along a line of sight
slightly out of the plane (solid curves), intersecting
the minor axis of the galaxy at two scaleheights
(maximizing \Eqt (\DTDM) for $S_{\rm l}$=0.5\rsun ; see \Sct 2).
The vertical dotted lines indicate the median radii for 
the \losa\ optical depth distributions.
All profiles are normalized to be 1 at maximum.  For NGC 3081, the
total \mlod\ (for a source at $R$=0 and $z$=2$h_{\rm z}$)
with ring is 1.3$\times$ that of only bulge and disk and
the mass in the ring is 20\% of the total mass.
For IC 4290, the corresponding figures are 1.4$\times$ and 22\%.
For NGC 7702, they are 1.5$\times$ and 25\%.
The dashed curves in the top and bottom panels repeat the
same for a three/four--exponential
bulge--disk fit to the surface--density profiles.
\label{GALS}
}
\end{figure*}

The high measured value of $\tau$ by Alcock \etal(1997) 
could be reproduced by a barred density distribution, 
if oriented almost parallel to the \losn\ (Zhao \& Mao 1996).
However, this is not within the scope of 
realistic models that reproduce observed stellar surface densities
and kinematics
(Binney, Gerhard \& Spergel 1997;  Bissantz \etal1997; 
Zhao, Rich \& Spergel 1996;
Nikolaev \& Weinberg 1997; Fux 1997; Sevenster \etal1999).
A larger scaleheight of the Bar and/or the Disk component, $\sim$400pc,
or scale of the inner Halo (see Minniti 1996) could also
increase the \mlod , but cannot be reconciled with the rapid decline
of \mlod\ with latitude (Alcock \etal1997). Hence,
a large apparent scaleheight is not an option (see also Sevenster \etal1999).
To the best of our present knowledge, 
the distribution of stars in the exponential Disk and the Bar or the Bulge
does not account for the observed \mlod , given the limits on
the mass--to--light ratio.

In this paper, we will try to resolve this controversy.
Throughout this article, we will adopt $R_{\odot}$$\equiv$ 8 kpc, unless
it is used as a generic symbol of distance between an observer
and the centre of a generic galaxy.

\section{How to enhance micro lensing ? }

The \mlod\ $\tau$ in a given direction depends on the
distance to the lenses, $S_{\rm l}$, and the sources, $S_{\rm s}$, via the 
term $ S_{\rm l} (S_{\rm s}-S_{\rm l})/S_{\rm s}$
(see Kiraga \& Paczynski 1994).
This is maximized, for one source--lens pair,
if $ S_{\rm l} = 0.5 S_{\rm s}$.
To increase the \mlod\ one would clearly 
like to add density roughly between the observer and the highest
density of sources: the central galactic--bulge region.
Let's define the contribution of the density at a distance $S_{\rm l}$
to the \mlod\ towards one source on the minor axis of the Galaxy,
at $S_{\rm s}\equiv R_{\odot}$ (the latitude of the source
has to be $<$ 8\degr\ so that $\cos(b)>$0.99), as

\eqnam\DTAU
 $$ {\Delta \tau (R, z )} =  { \rho(R,z) }
     { {S_{\rm l}(R_{\odot}-S_{\rm l})} \over {R_{\odot}}  } \Delta R$$

 $$ \ \ \ \ \ \ \ \ \ \ \ \ \ \ \
       = {\rho_0(R) \exp[-S_{\rm l} \tan |b| / h_{\rm z}]}
      { {S_{\rm l}(R_{\odot}-S_{\rm l})} \over {R_{\odot}} }\Delta R,
\eqno(\new)  $$

\noindent
with $(R,z)$ galacto--centric cylindrical coordinates and $b$ galactic
latitude,
and the total mass involved in creating that density at that
distance as

\eqnam\DMAS
 $$ \Delta M (R) = 4 \pi h_{\rm z} R \rho_0(R)\Delta R
       = 4 \pi h_{\rm z} (R_{\odot}-S_{\rm l}) \rho_0(R)\Delta R ,
\eqno(\new)  $$

\noindent
assuming the density is completely axisymmetric with 
constant exponential scaleheight $h_{\rm z}$.
The quantity one would like to maximize is

\eqnam\DTDM
 $$ \Delta \tau / \Delta M = { {S_{\rm l}} \over {4 \pi h_{\rm z} R_{\odot}}  }
   \exp[-S_{\rm l} \tan |b| / h_{\rm z}]  ,
\eqno(\new)  $$

\noindent
which is maximal for $S_{\rm l}$=$R_{\odot}$ if $b$=$0^{\circ}$
and for $S_{\rm l}$=$h_{\rm z} / \tan |b|$ otherwise.
(Beware that \Eqt (\DTDM ) only optimizes for fixed $\tan |b|$;
it is not a true free parameter.)
In the Galaxy, with $h_{\rm z}$$\sim$300 pc and the microlensing
measurement at $b$$\sim$4\degr , this
gives $S_{\rm l}$=4.3 kpc (or galacto--centric radius $R$=3.7 kpc).

Would it be an option to have a ring of enhanced density at such 
a radius ? In barred galaxies, one commonly observes dense rings,
that trace closed orbits near resonances in the rotating
potential (see Buta \& Combes 1996 for a review).
They contain significant mass.
A so--called ``inner ring'', encircling
the galactic Bar, may prove an ideal candidate.

\fignam\HIYY
\anfig
\vskip .5truecm
{\psfig{figure=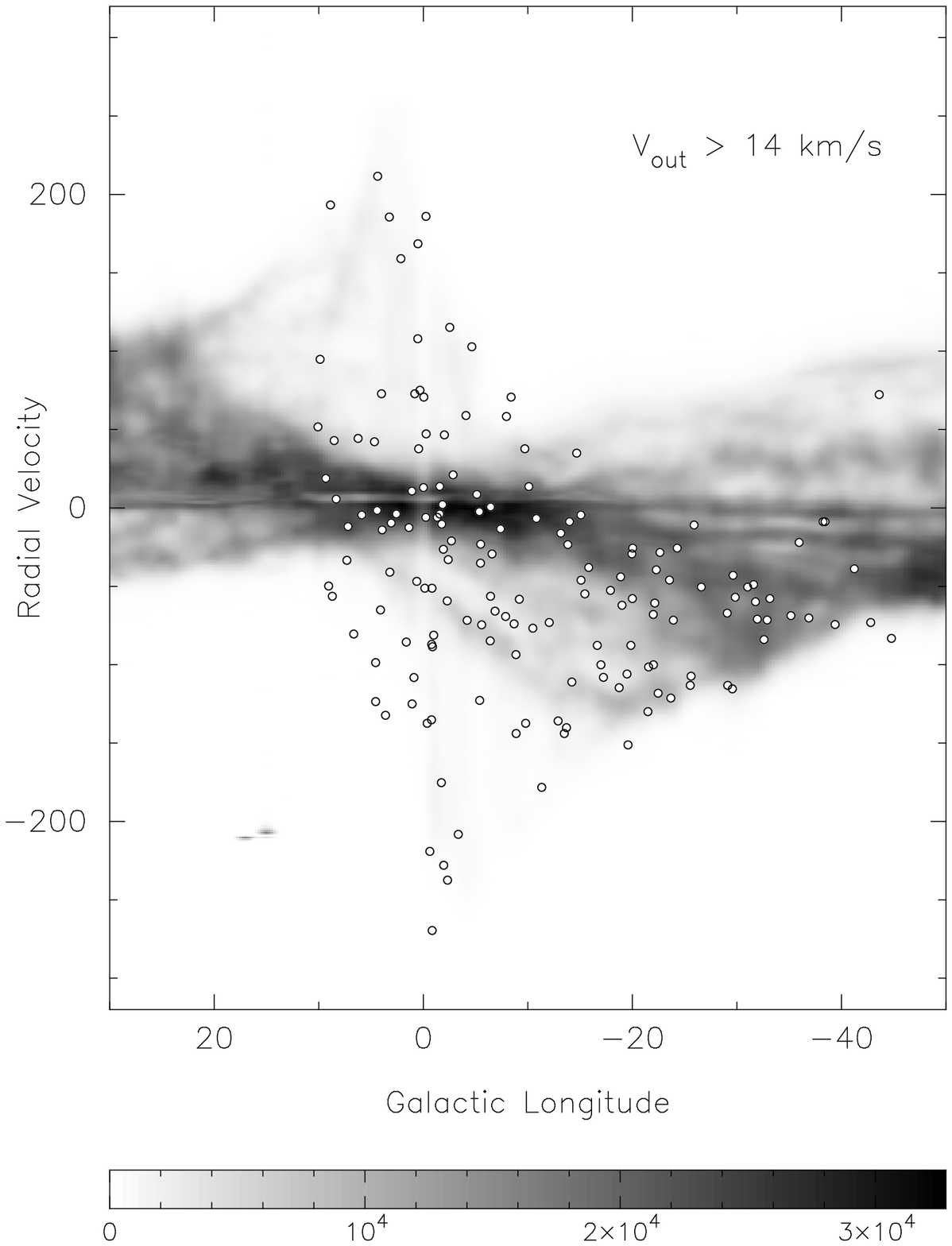,width=7truecm}}
\figcaption{
The HI \lvd\ (greyscale) for latitudes $|b|$$<$\decdeg1.5,
with higher--mass OH/IR stars overplotted (circles; see \Sct 3).
The ``3--kpc arm'' is the filament extending from 
($\ell,V$)$\sim$($-$20\degr ,$-$130 \kms ) 
to $\sim$(+5\degr ,$-$25\kms ). A suggested
counterpart, the ``4--kpc arm''
(see Shane 1972), turns around at $\sim$(+25\degr , 100\kms ).
\label{HIYY}
}
\vskip .5truecm

We demonstrate this in \Fg\GALS , with
surface--density profiles, in the reddest available bands,
and the corresponding $\Delta\tau$ (\Eqt\DTAU ) for 
three external inner--ring galaxies, 
NGC 3081 (SAB), IC 4290 (SB) and NGC 7702 (SA).
We assume a constant, exponential scaleheight, axisymmetry
and constant mass--to--light ratio across all galaxies and 
ignore inclination effects ($i$$>$$40^{\circ}$ for all;
inclination smears out density contrasts, but does not change 
the relative masses of components).
The spatial density in the plane of the galaxies
hence follows the surface--density profiles.
We place the observer at exactly twice the ring radius in 
each galaxy.
This optimizes \Eqt(\DTDM ) for $\tan |b| = 2h_{\rm z}/R_{\odot}$, 
as is clear from the differences between the
two lines of sight through the mass models in \Fg\GALS .

In all three galaxies, lensing material well in front of the bulge 
yields 50\% of the \mlod\ toward a source situated at the centre,
compared to about 30\% at the same radii
for the fits without ring (see \Fg\GALS ).
For the \losn\ out of the plane, the rings themselves
contribute 30\% to 50\% of the \mlod\ due to bulge and disk alone.
For a lower $\tan |b|$, assuming smaller scaleheight while
keeping the ring halfway between observer and centre of the
galaxy, the relative contribution is still larger.
For NGC 7702, for instance, the ring contribution is 60\% when
assuming a scaleheight that is half as large as used in \Fg\GALS .

For simplicity and clarity, in this section we 
calculated \mlod\ contributions with
just one source on the galaxy's minor axis. Naturally,
for the full \mlod\ one
needs to take into account the full distribution of sources.
For source--detectability parameter $\beta$=0 (see \Sct 4)
and $S_{\rm s} < R_{\odot}$, the rings' relative contributions
to the true \mlod\ are virtually the same as the fractions given 
in the caption of \Fg\GALS , confirming the validity of our simpler
approach. For smaller distance cut--off the fraction is higher,
and vice versa; for $\beta$=$-$1 the fraction is lower.

\section{An inner ring in the Galaxy}

The ``3--kpc arm'' is a prominent large--scale feature seen in
the kinematic distribution of galactic hydrogen 
(see Oort(1977) for early references). The HI 
emission can 
be followed over a range of longitudes , from $\ell$$\sim$7\degr\ to 
$\ell$$\sim$$-$20\degr\ (\Fg\HIYY ). 
In front of the \gc\ it is seen in absorption at a \losa\ velocity 
of $-$53\kms . This measurement gives one of the most
convincing arguments for non--circular motions in the galactic Disk.
Many explanations have been proposed for the
origin of the arm, ranging from an explosive event in the \gc\ to
non--circular motions forced by a rotating bar
(Shane 1972; Yuan 1984; Binney \etal1997; 
Englmaier \& Gerhard 1999).

In most stellar samples such a longitude--velocity structure 
cannot be traced, for obvious reasons.
However, we can identify nine OH/IR stars from
a systematic sample 
as ``3--kpc'' stars (\Fg \HIYY ; see Sevenster 1999 for details). 
They do not only share their
longitude--velocity structure with the HI filament, but also
their latitudes : all are well within a degree from the plane.
This is the first unambiguous evidence for a stellar--kinematic counterpart
of the 3--kpc arm. The initial masses of these
nine objects are $\sim$4\msol , corresponding to ages of $\sim$0.2 Gyr.
Of the older OH/IR stars of the same sample, only one coincides
with the 3--kpc longitude--velocity filament. 
The probability for a chance alignment with 
these properties is less than 1\%.

For two of those stars (IRAS 17378$-$3100 and 17271$-$3425), distances
are given in the literature: 4.9 kpc and 4.3 kpc, respectively
(L\'epine \etal1995).  Although these fit in perfectly
with our scenario, they are derived using, amongst others,
[K$-$L], which may vary by $\pm$1 mag between
stellar maximum and minimum (as for 17271$-$3425)
resulting in a $\pm$2 kpc error for this source from the
colour variability alone. This distance information clearly
does not weaken our arguments, but neither does it strengthen them.

\begin{figure*}
\fignam\BWF
\anfig
{\psfig{figure=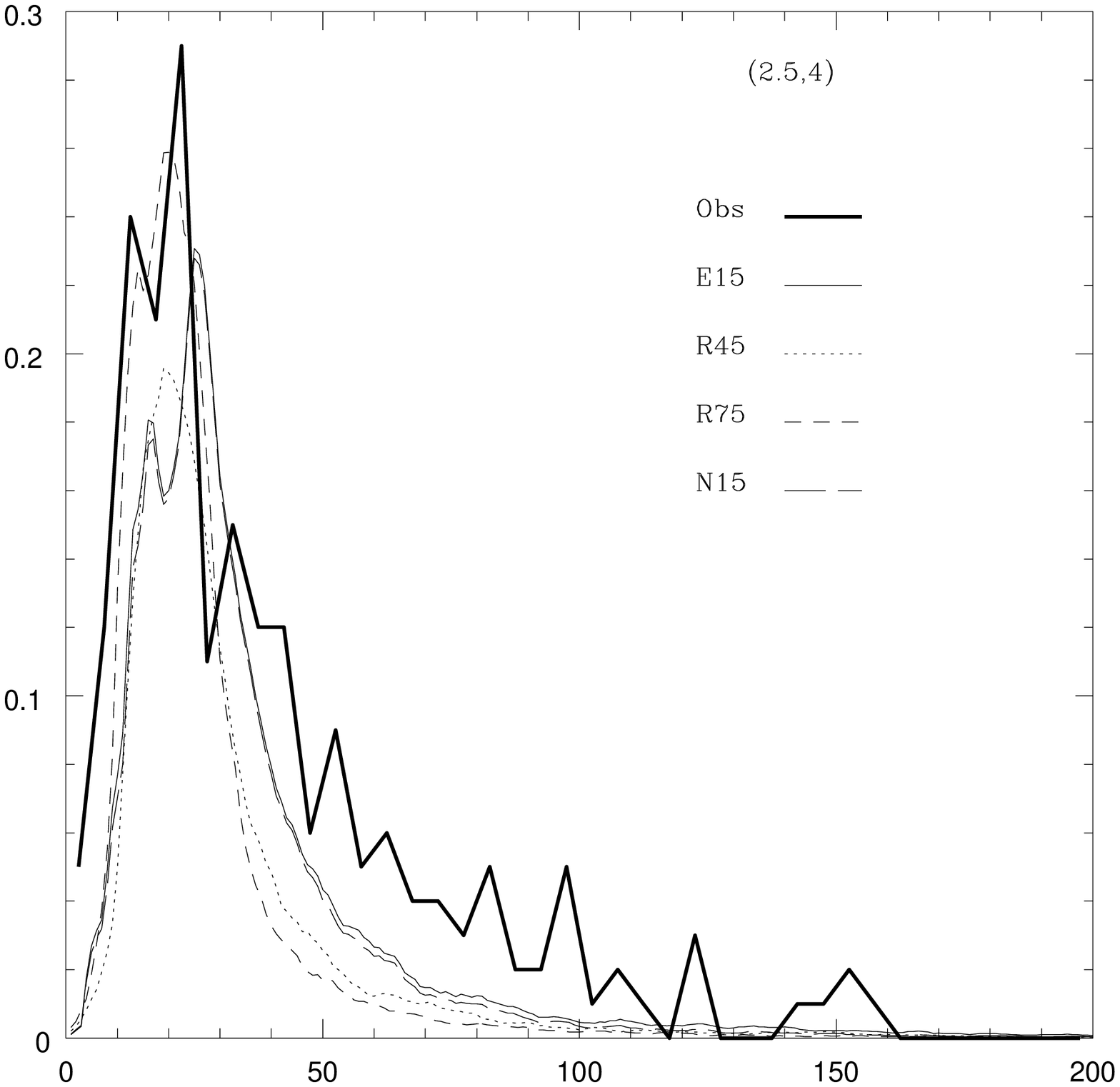,height=7truecm}}
\vskip -7truecm
\hskip 7truecm
{\psfig{figure=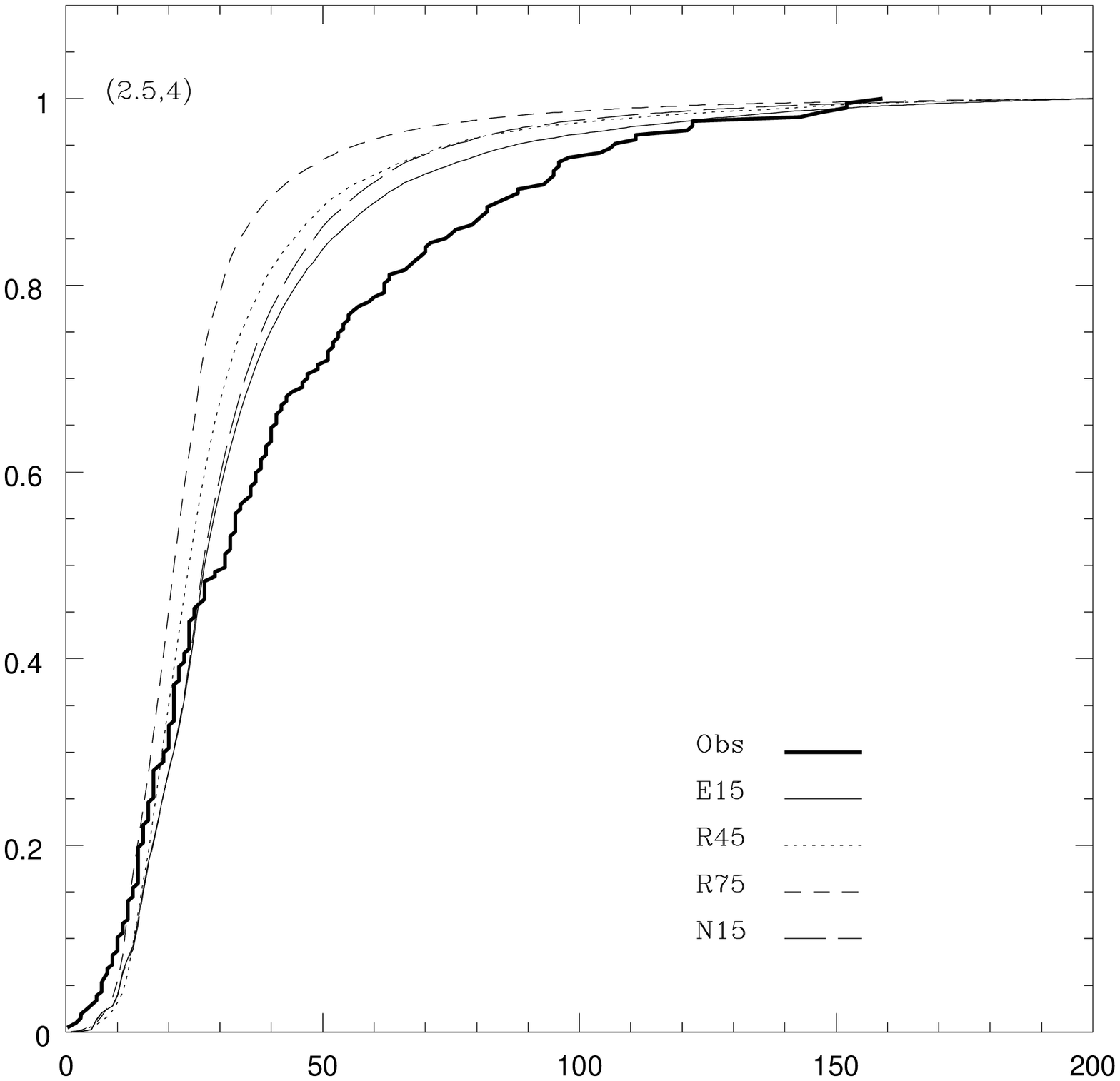,height=7truecm}}
\figcaption{
The relative (left) and cumulative relative (right)
event--duration distributions
($\beta=-1,m\equiv1$) along the line of sight toward Baade's window
for four different models and the MACHO observations (thick curve;
202 events from the MACHO alert webpage (1995--1999,
well--determined durations $<$ 200 days only, excluding field 301).
Bootstrapping the model distributions, we find that
there are no biases in the CEDDs and
the errors are below 0.01 (largest around 20 days).
Note that the models are not fits to the observations.
The ordinate gives the event duration in days.
\label{BWF}
}
\end{figure*}

Had the 3--kpc arm been a 
density wave, the stars and the gas would be likely to 
have separated during 0.2 Gyr.
Now, they are separating only on timescales expected from normal
diffusion ($\Delta U \sim$10 \kms\ for ages of $\sim$ 0.25 Gyr,
Wielen 1977).
The ages of the stars and the 53\kms\ expansion rate also rule out an
explosive origin of the 3--kpc arm. They should be at a radius
of 10 kpc by now, whereas the current position is
about half way between the Sun and the Galactic center.     

The evident explanation for this close association of
the OH/IR stars and the hydrogen, that has survived for 
2 galactic revolutions, is that they move along the same 
single filled elliptical orbit, swept up by the rotating Bar 
into an inner ring (\Sct 2).
This is in agreement with independent determinations
of the pattern speed (eg. Kalnajs 1991; Kalnajs 1996; Fux 1997).
The negative--longitude tangent point is
clearly identified at $-$22\deg\ (Shane 1972; Hammersley \etal1994; Sevenster
1999). On the positive--longitude side, the tangent point
is likely to be around +25\degr $\pm$2\degr\ 
(Shane 1972; 
Cohen \etal1980; 
Hammersley \etal1994; 
Blommaert, van Langevelde \& Michiels 1994). 
Since inner rings share the major axis with the bar they 
encircle, as corroborated by the tangent point at a radius of 3 kpc,
it is even less likely that we see the Bar nearly end--on (see \Sct 1).
For small viewing angles, the velocity would vary much more slowly 
with longitude than observed.

Based on external ringed galaxies (\Sct 2), 
the 3--kpc ring should be a good candidate to reproduce the observed
value of the \mlod , in the same fashion as the galaxies in \Fg\GALS .
In the next sections, we will discuss some implications of this
proposition, while avoiding to make absolute statements, as any choice
of values of parameters for the ring or the other galactic components
would make the discussion less general. 

\section{Microlensing measurements}

For a sample of 45 MACHO events, the calculated
\mlod\ is $2.11^{+0.47}_{-0.39}\times 10^{-6}$,
roughly corrected for blending, toward
($\ell$ = \decdeg 2.7, $ b=-$\decdeg4.1) on average
(Alcock \etal1997). A more recent measurement, using
difference--image analysis (Alcock \etal2000a), is even higher, 
$2.43^{+0.39}_{-0.38}\times 10^{-6}$ toward 
($\ell$ = \decdeg 2.68, $ b=-$\decdeg3.35).
Fewer uncertain factors are involved when using
red--clump--giant events only,
although unfortunately the number statistics are
not very favourable for that sample; Alcock \etal(1997) give
$\tau$=$3.9^{+1.8}_{-1.2}\times 10^{-6}$
toward ($\ell$ = \decdeg 2.55, $ b=-$\decdeg3.64),
but Popowski \etal(2000) give $2.0^{+0.4}_{-0.4}\times 10^{-6}$
toward ($\ell$ = \decdeg 3.9, $ b=-$\decdeg3.8) from a
preliminary analysis of the clump giants in the full MACHO 
sample, using four times more sources than in 
the Alcock \etal(1997) calculation.

\begin{figure*}
\fignam\FTE
\anfig
{\psfig{figure=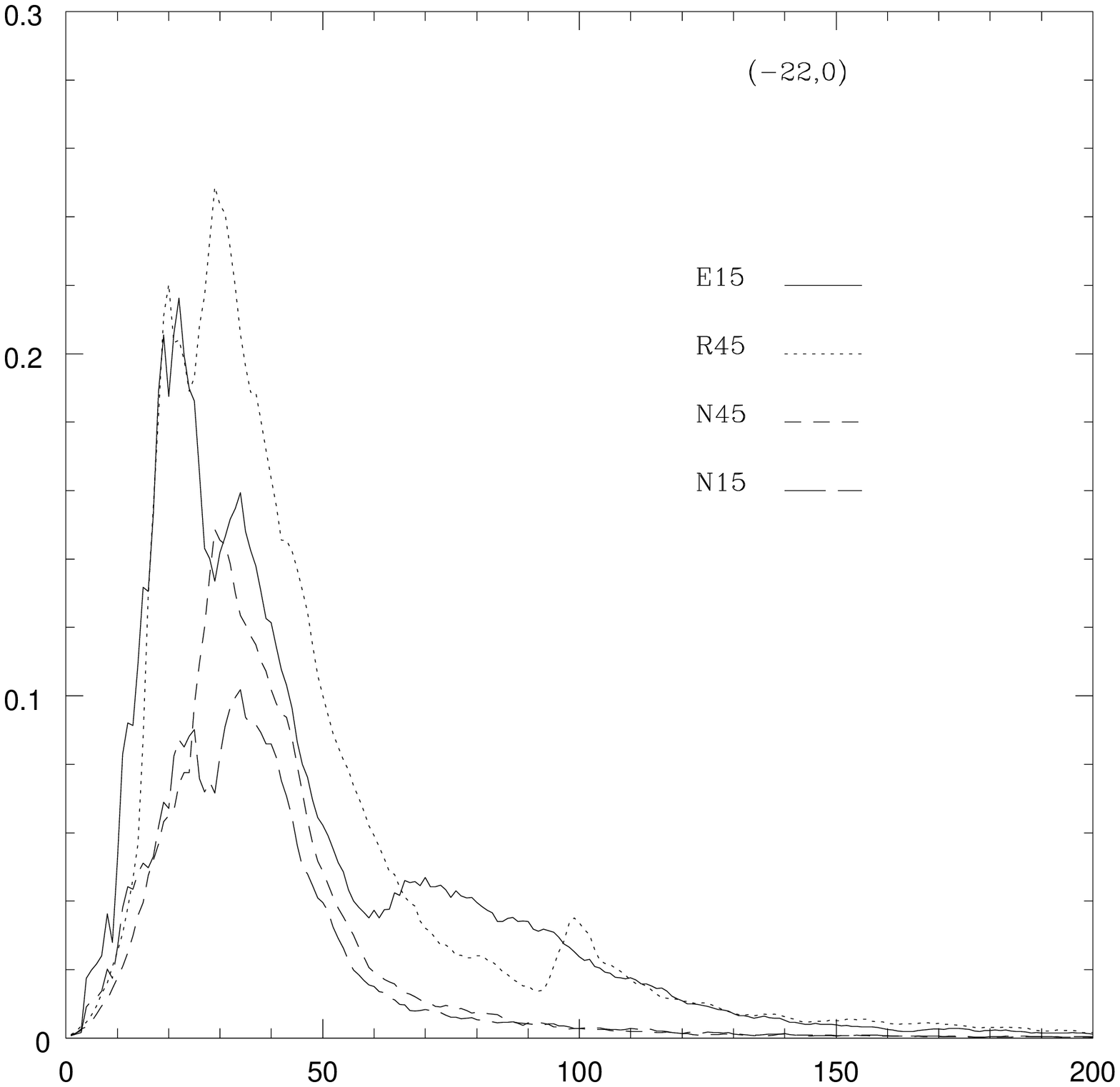,height=6truecm}}
\vskip -6truecm
\hskip 5.2truecm
{\psfig{figure=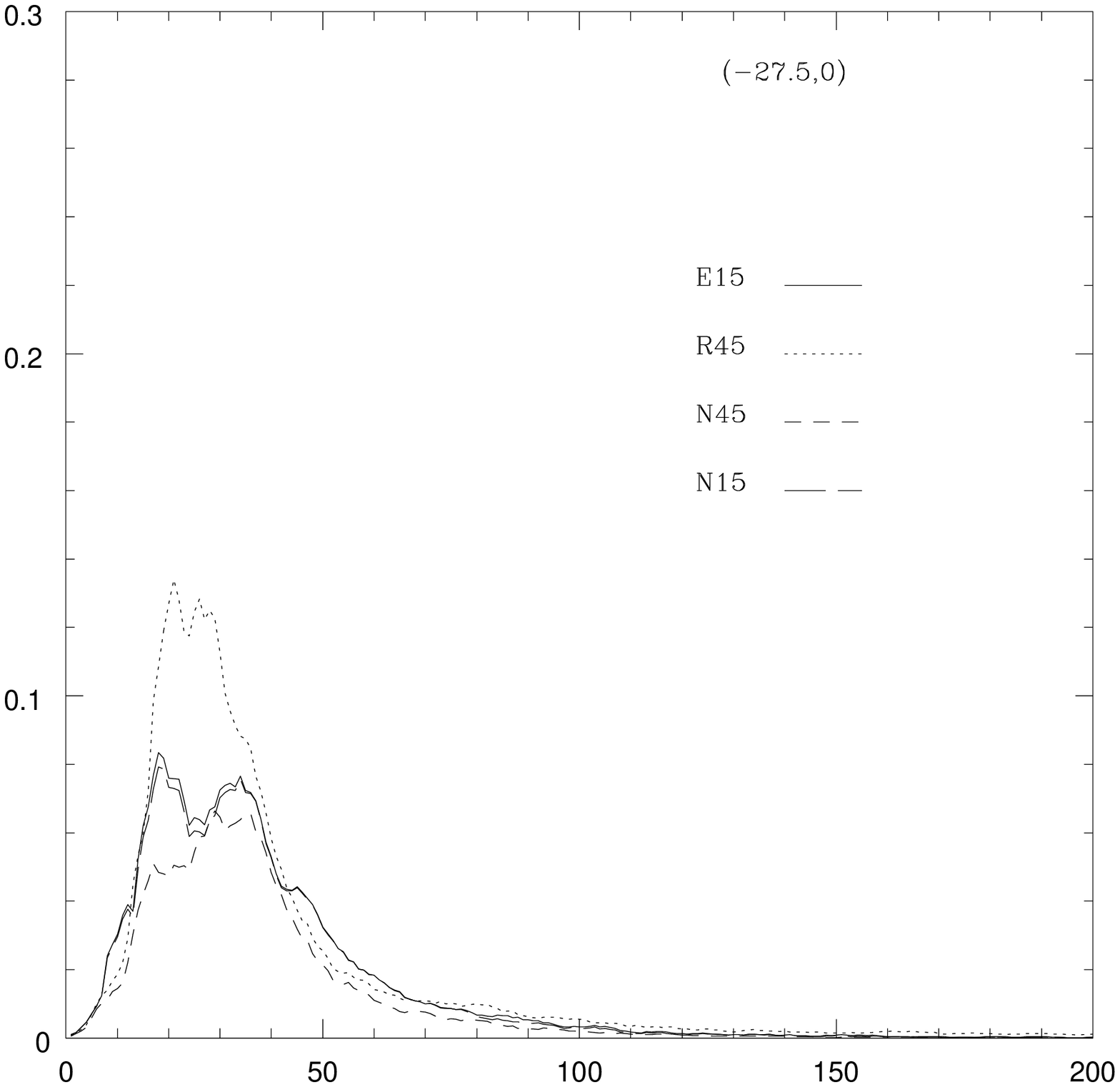,height=6truecm}}
\vskip -6truecm
\hskip 11truecm
{\psfig{figure=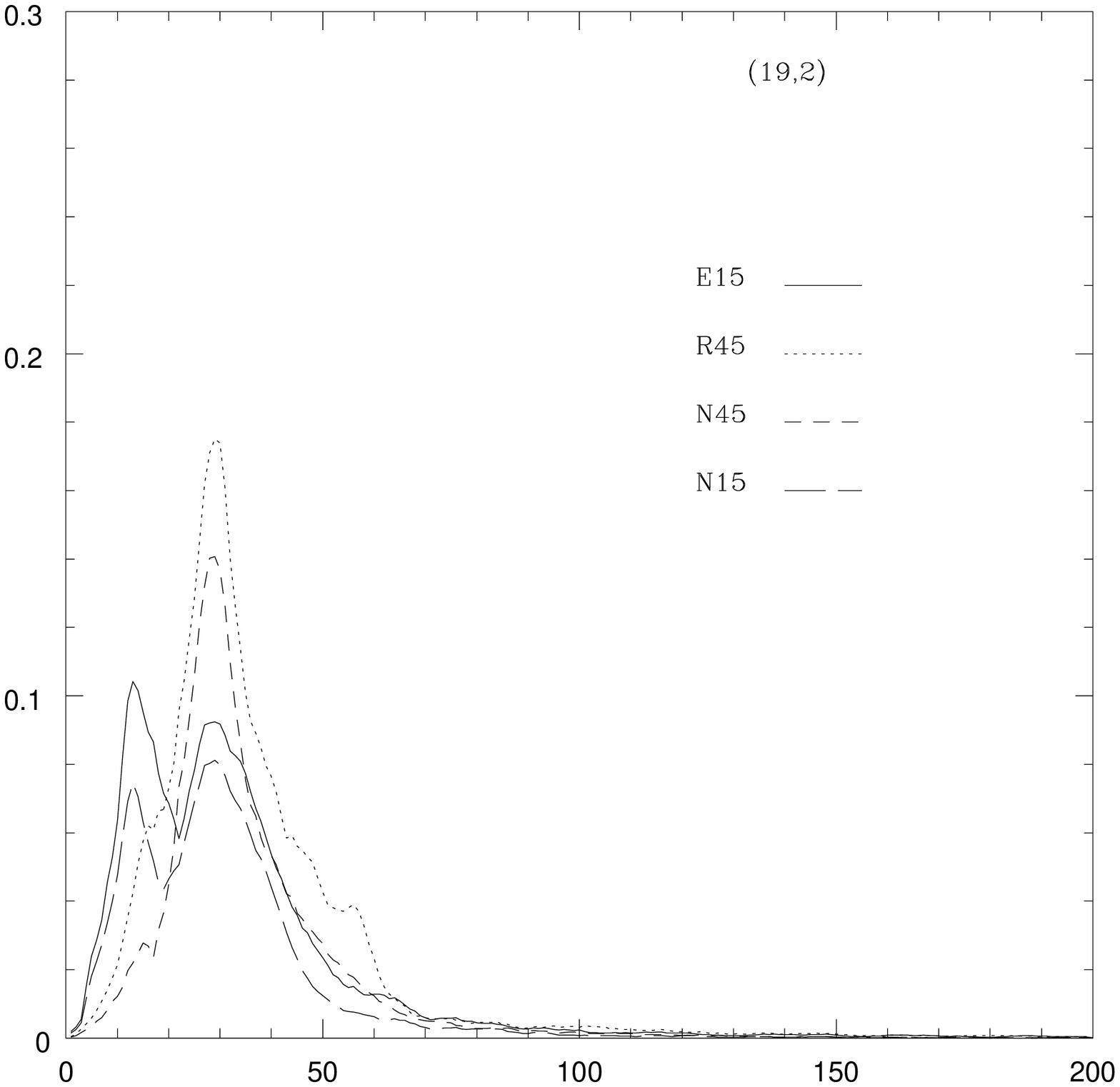,height=6truecm}}
\figcaption{
As \Fg\BWF a, along lines of sight toward 
the negative tangent point (a), a disk EROS field (b) and
the high--longitude MACHO fields (c) for the models
E15 and R45 and their counterparts without ring particles (N15,N45).
The longitude and latitude in degrees of the lines of sight
are given between brackets in the top right of each panel.
The vertical scale is arbitrary, but the same for all 
panels (as well as \Fg\BWF a). 
The ordinate gives the event duration in days.
}
\end{figure*}

When modelling the \mlod , an important parameter is 
the power of the source--distance dependence, $\beta$ 
(see Kiraga \& Paczynski 1994).
In the full expression for $\tau_{\beta}$ the source 
distance comes in as $S_{\rm s}^{2+2\beta}$, so that 
for $\beta$=$-$1 the number of detectable sources goes up with
$S_{\rm s}^{2}$, as it should from the usual geometrical argument,
and down with $S_{\rm s}^{-2}$, as the sources get fainter.
For sources with a very narrow range of luminosities and 
distances, such as red--clump giants that are very bright and
only in the bulge, $\beta$=0 is more appropriate since they
are all bright enough to be detected.

Importantly, the observations were not
taken exactly towards the \gc . This makes
the case for the inner ring stronger, as we already saw in \Sct 2.
For the latitude of the observations, the 
maximum of \Eqt(\DTDM ) would occur 
exactly half--way between the Sun and the \gc\ for a scaleheight 
of 280 pc.
Typical model values for the \mlod\ are $0.9 \times 10^{-6}$ for $\beta$=$-$1 
and $1.4 \times 10^{-6}$ for $\beta$=0 (Bissantz \etal1997; 
Sevenster \etal1999). A ring 
contribution of $\sim$50\% of the bulge--disk contribution (\Sct 2) 
would bring models
to within 1.5$\sigma$ below the Alcock \etal(1997) values
and $\beta$=0 models right on the Popowski \etal(2000) value.
For comparison, the 
spiral--arm model of Taylor \& Cordes (1993), for example,
would yield only of the order of a few per cent of the bulge--disk
\mlod\ (and that is assuming
the spiral--arm mass contrast is the same for stars as it
is for gas).

The ring's contribution, in depending on the distance to 
the observer, will depend crucially on the semi--major axis as well
as the viewing angle of the ring. Other factors are the radial
and vertical extent of the ring, as discussed extensively in \Sct 2.
Note that \Eqt (\DTDM) is particularly suited to approximate the
clump--giant measurement, with the sources close to the Galaxy's 
minor axis and detection chance $\equiv$1.

\begin{figure*}
\fignam\CFTE
\anfig
{\psfig{figure=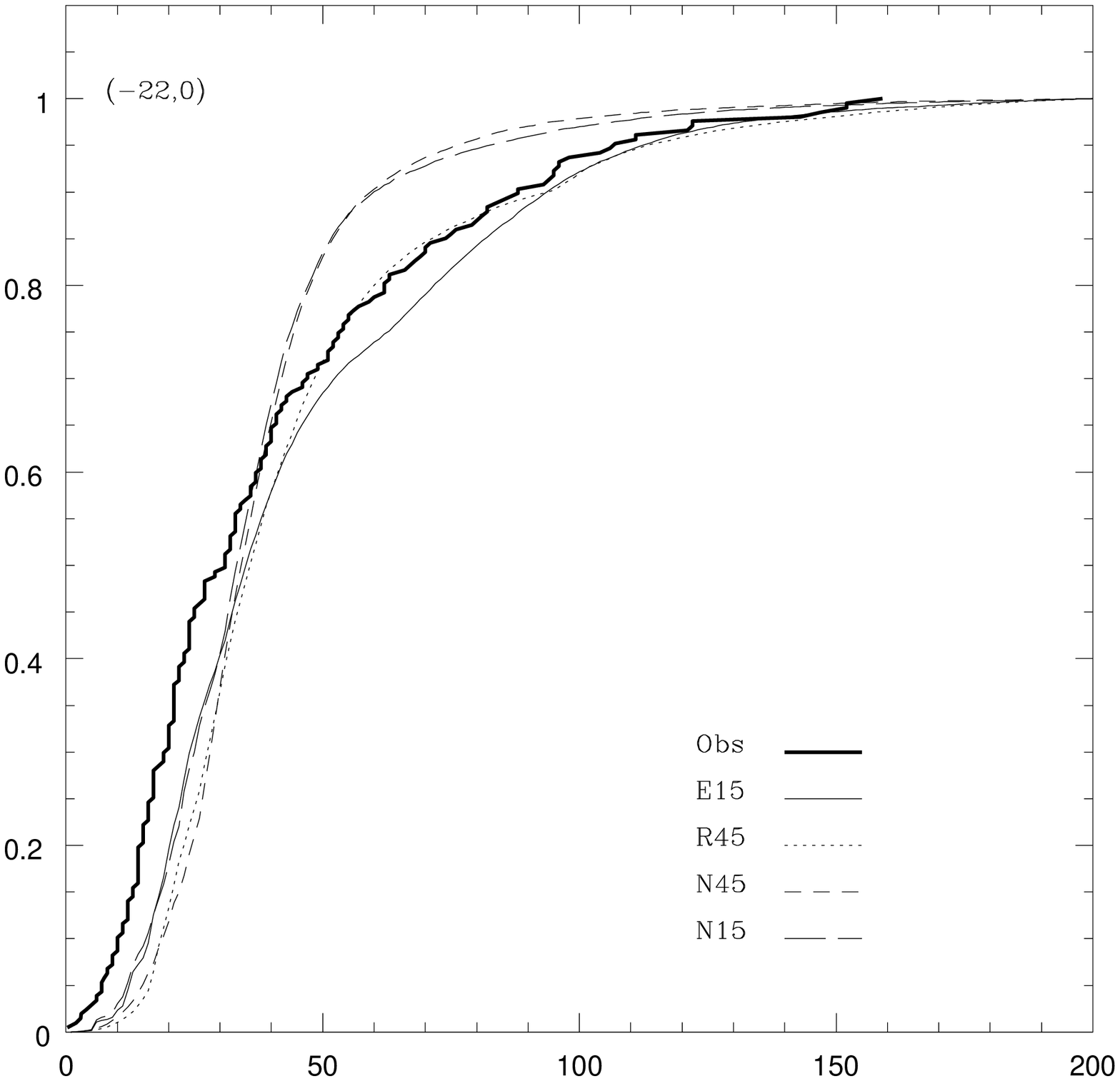,height=6truecm}}
\vskip -6truecm
\hskip 5.2truecm
{\psfig{figure=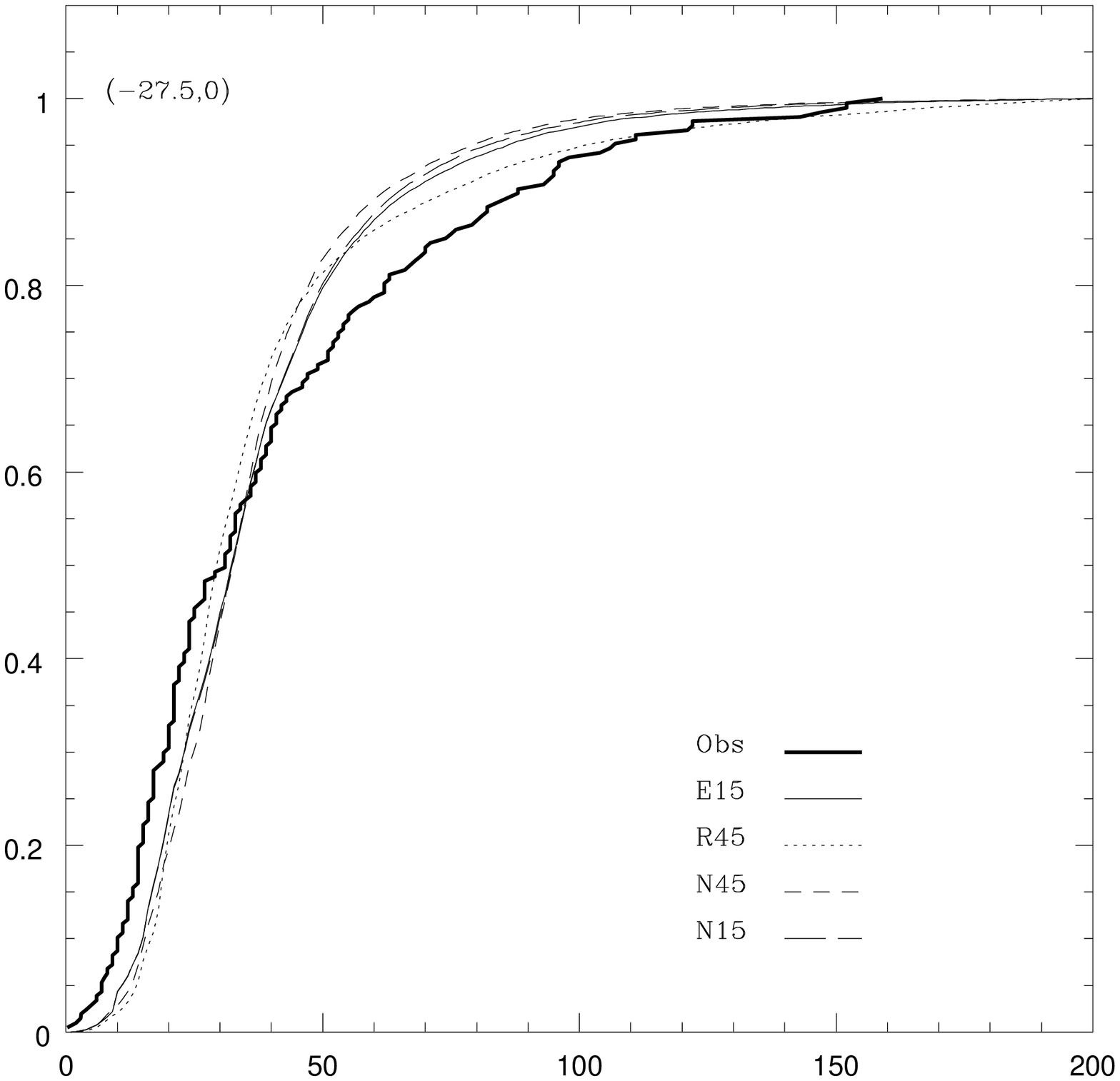,height=6truecm}}
\vskip -6truecm
\hskip 11truecm
{\psfig{figure=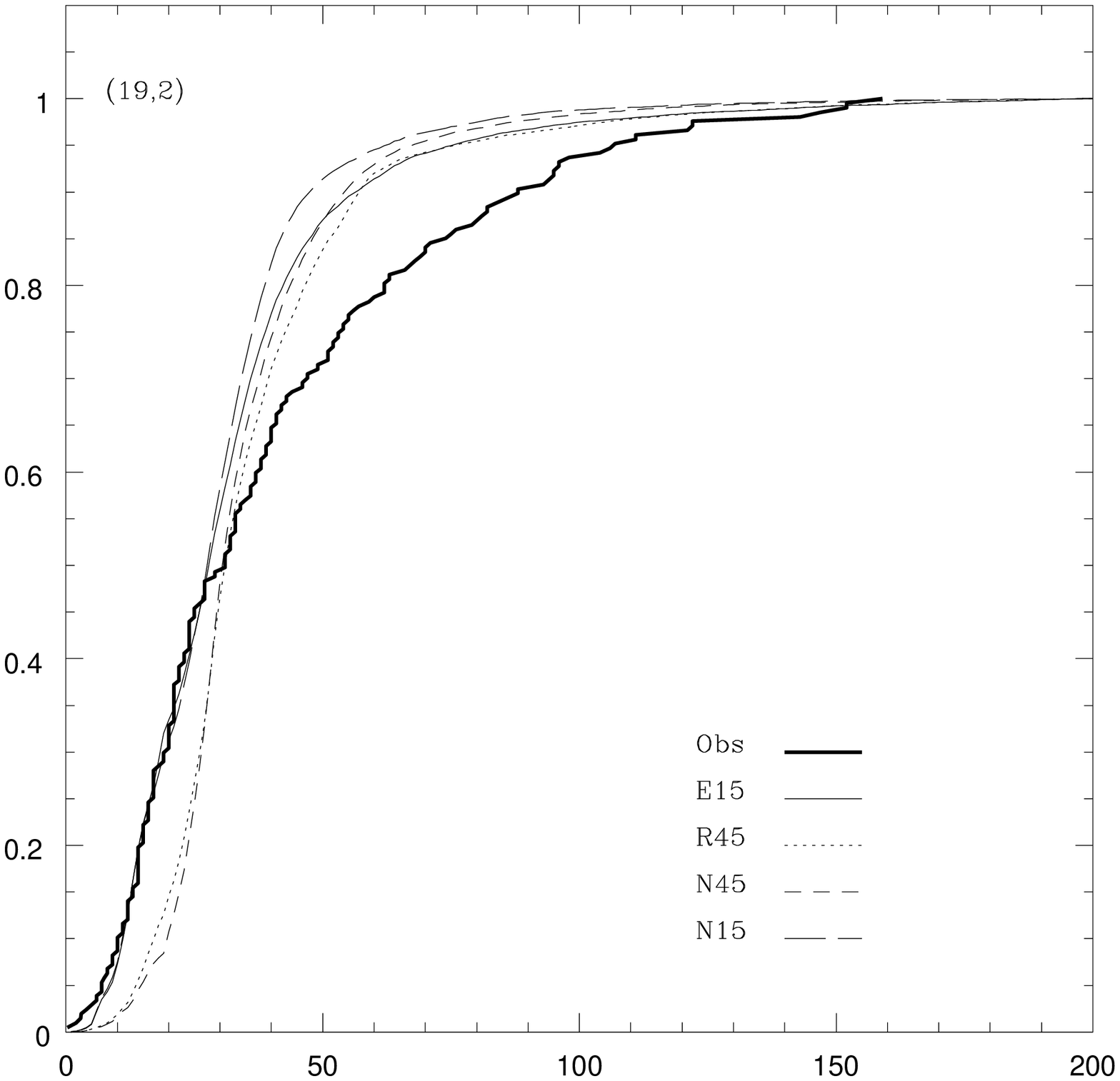,height=6truecm}}
\figcaption{
As \Fg\FTE , for the cumulative density distribution
of event durations. The observed distribution toward
Baade's window is only
shown for comparison with \Fg\BWF b.
}
\end{figure*}

\section{Event--duration distributions}

\subsection{Models}

For the remainder of this paper, 
we will concentrate on the expected influence of a
ring of lenses between us and the \gc\ on 
the event--duration distribution of microlensing events, 
relative to the durations  of events in an underlying generic
bulge--disk distribution.
For this, we will need to know distances, positions and transverse
velocities -- in other words, almost the full six--dimensional
coordinates -- for sources and lenses, in the ring
as well as the disk and bulge.
We use an N--body simulation of the barred Galaxy by Fux (1997);
model ``m08t3200'', with $R_{\odot}$=8 kpc,
$V_{\odot}$=210 \kms , \fv =1 and adaptable $\phi_{\rm bar}$
(for a justification see also Sevenster \etal1999).
A random realization of the following ring model is added,
assuming that all ring particles stream strictly along the
ring--density contour :

\eqnam\RHOR
$$ \rho_{\rm ring} = \rho_{\rm 0r}
     \exp(-(R_{\rm r} - a )^4/(2\delta^4_{\rm r}) )
       \exp(-|z|/h_{\rm z}) ,  $$
$$\ \ \ \ {\rm where}\ R_{\rm r}^2=x^2_{\phi}/q^2_{\rm r}+y^2_{\phi}\ {\rm and}  $$
$$\ \ \ \ x_{\phi},y_{\phi}\rm\ carthesian\ minor/major\ axis\ coordinates .
\eqno(\new)  $$

\noindent
As we argued in \Sct 3, the ring is a filled elliptical orbit, so
the velocities of particles on the ring are derived
assuming angular momentum is conserved. This is not strictly
true in a non--axisymmetric potential, but a good approximation.
Thus, the magnitudes of the ring velocities are found by fixing
the angular momentum to give a radial velocity of to +53 \kms\ at
the position where the ring intersects the
\losn\ toward the \gc\ (with $V_{\odot}$=210\kms , $U_{\odot}$=0\kms ). 

The fractional width of the inner ring in the minor--axis
H--band profile of eg.~NGC 3081 is about 0.2 and data on
other galaxies suggest that this is a fairly standard value.
Hence, we derive that the velocity dispersion in the galactic ring
would be $<$0.25*200 \kms = 50 \kms .
In the following text, we will mention the influence of such
a velocity dispersion in the several cases that will be discussed.
However, since we are mainly interested in the
influence of the geometry of the ring -- ie.~the streaming velocity -- the
standard models are without velocity dispersion. This in fact
turns out to yield perfectly acceptable approximations.

The total number of particles
in the ring is taken to be 10\% of the number in the N--body model,
$\delta_r=600 \rm pc$ and $h_z$=380 pc. 
(This scaleheight equals that of the N--body model (see Fux 1999)
and maximizes \Eqt(\DTDM ) for $R$=2.6 kpc).
A range of $a,\phi,q$ are used (Table \PARS).
The ``Nxx'' runs have no ring, only $\phi_{\rm bar}$=xx.

\subsection {Results and comparison to available data}

The event--duration distribution (EDD) in the direction of the
Bulge (all fields with $ 0^{\circ} < \ell < 10^{\circ}$), for 
true micro--lensing events with a well--determined duration
(\^t , Alcock \etal1997), shows
a peak at 15--20 days with a fairly massive tail reaching well up to
150 days (data from the MACHO alert webpage
(http://darkstar.astro.washington.edu), hence not 
complete, but the distribution is similar to
that in Alcock \etal1997). The median duration for the 202 events
is 29 days (see \Fg\BWF ).
Models with a range of Bulge and Disk parameters and a range
of forms of the initial--mass function (IMF) fail to reproduce
the observed EDD consistently. The short--event tail may
be explained by changing the lower mass cut--off
and the exponent of the IMF, or simply the blending correction.
The long--event tail of the observed distribution
is more difficult to reproduce, but potentially very important. 
Even though it consists of only four
events (\^t $>$ 100 days) in the fully corrected sample
(or 10 in the sample of \Fg\BWF ),
these do make up about one third of the total \mlod\ toward Baade's 
window (Alcock \etal1997).

From the relative proper motions and distances of our model
(\Sct 5.1), we calculate the EDD for
several lines of sight, following Han \& Gould (1996).
We set all masses to 1, to separate the influence on the \mlod\ of the 
kinematical distribution from that of the initial--mass function.
This means we do not try to reproduce the observed MACHO EDD.
In \Fg \BWF ,\FTE \&\CFTE, we show (cumulative) event--duration distributions 
((C)EDD) for several models (Table \PARS), 
along several lines of sight. We discuss
those in detail in the following subsections, but note that all
models, with and without ring,
with a viewing angle of 15\degr\ give bimodal distributions 
for \^t $<$50 days for all lines of sight.

\subsubsection{Toward the far tangent point}

From the schematic drawing 
in \Fg \STRM\ it is evident that near the far tangent point,
a ring will give rise to a considerably different distribution of
transverse velocities than a spiral arm.
For a spiral arm, the transverse
velocities will be slightly larger than for the underlying disk
around the distance of the
tangent point and slightly smaller just in front of and behind it.
This would not create an EDD significantly different
from that of the disk.
However, for a ring, the transverse velocities 
will be much smaller than in the underlying disk and, for certain
viewing angles, very similar
for a large range of distances along the \losn .

\fignam\STRM
\anfig
\vskip .5truecm
{\psfig{figure=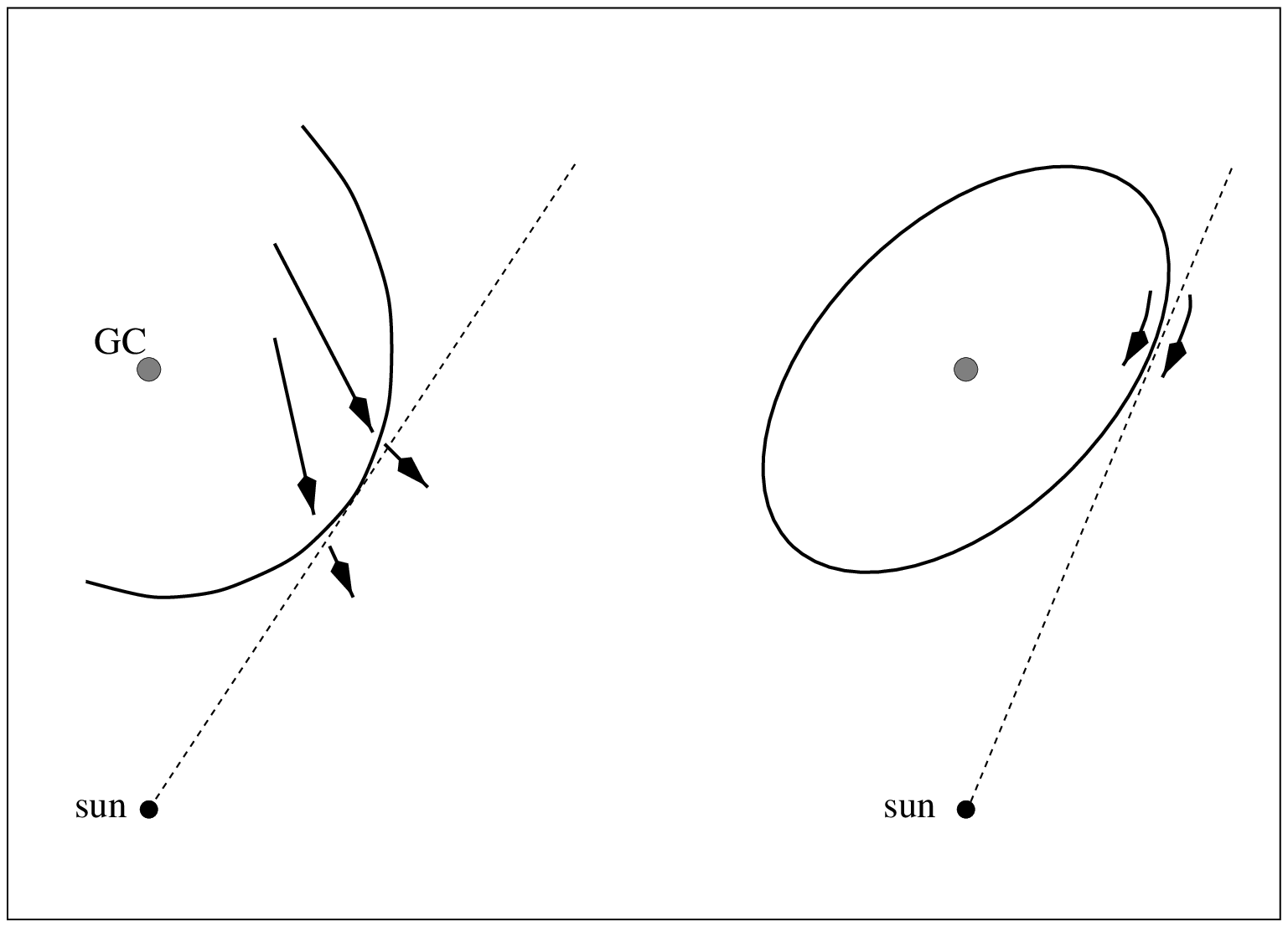,width=8truecm}}
\figcaption{
For a ring, the average stellar velocities will be along the \losn\ at
the tangent point, as opposed to the situation for a spiral arm.
Therefore, the tangential velocities will be negligible for a ring,
and will be coherent over a relatively large distance range.
A spiral scatters stars passing through it onto more radial
orbits, without necessarily making the tangential velocities more coherent. 
We can therefore expect that {\it at the tangent
point}, a ring will give longer events on average than a spiral.
}
\vskip .5truecm

\tabnam\PARS
\begin{deluxetable}{crrrrrrrrrrr}
\tablecaption{ Parameters for the runs.}
\tablewidth{0pt}
\tablehead{
\colhead{\bf Name } &
\colhead{$\phi$ } &
\colhead{a } &
\colhead{q } 
}
\startdata
   &$^{\circ}$& kpc &  \\
{\bf E15,30}& 15,30 & 5 & 0.6 \\
{\bf R15,30}& 15,30 & 4 & 0.8 \\
{\bf E45,75}& 45,75 & 5 & 0.6 \\
{\bf R45,75}& 45,75 & 4 & 0.8 \\
\enddata
\tablecomments{
For all runs, $\phi_{\rm bar} \equiv \phi_{\rm ring}$ (\Sct 5.1).}
\end{deluxetable}

\begin{figure*}
\fignam\DMLR
\anfig
{\psfig{figure=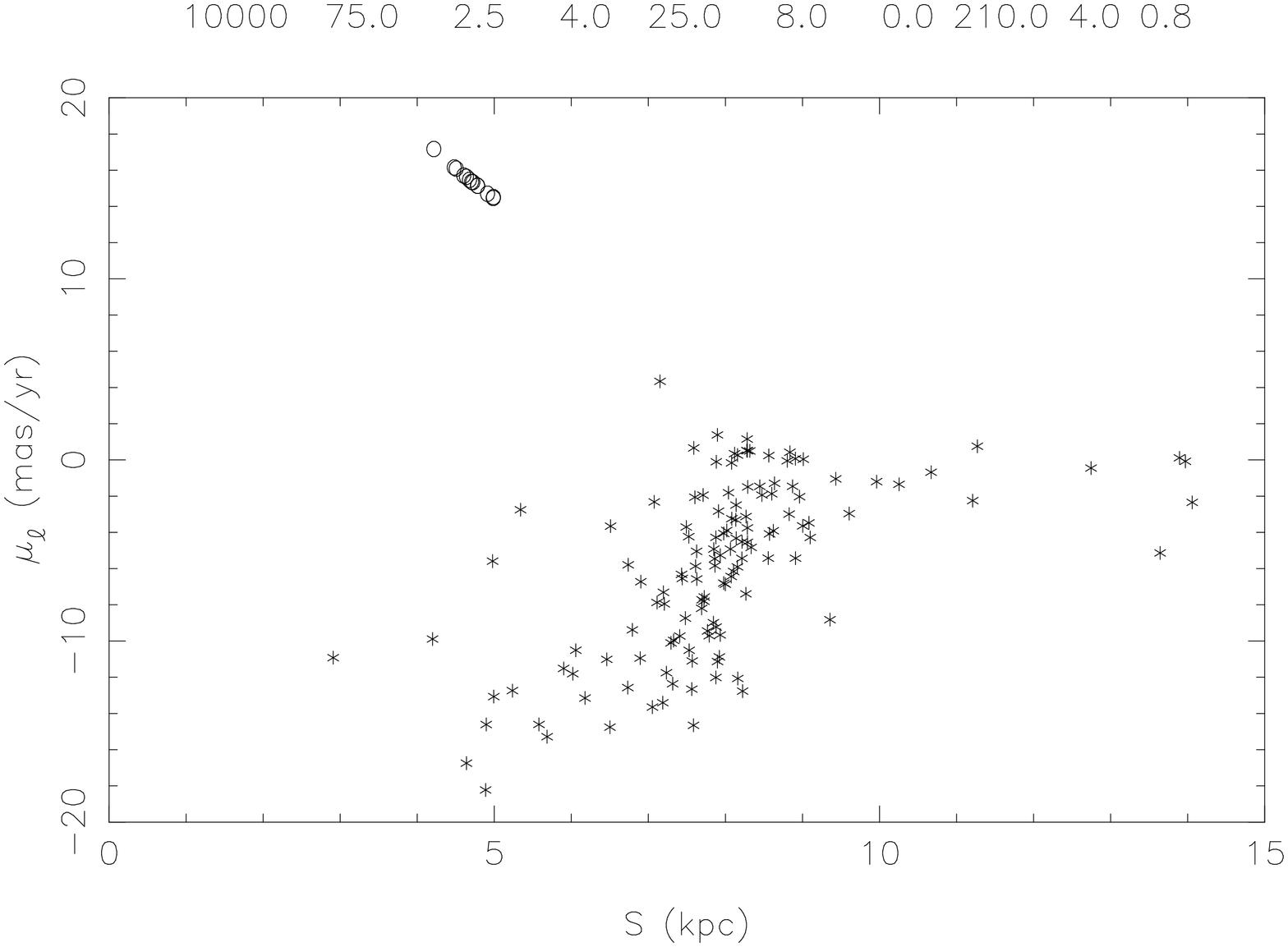,angle=270,height=6truecm}}
\vskip -6truecm
\hskip 7.5truecm
{\psfig{figure=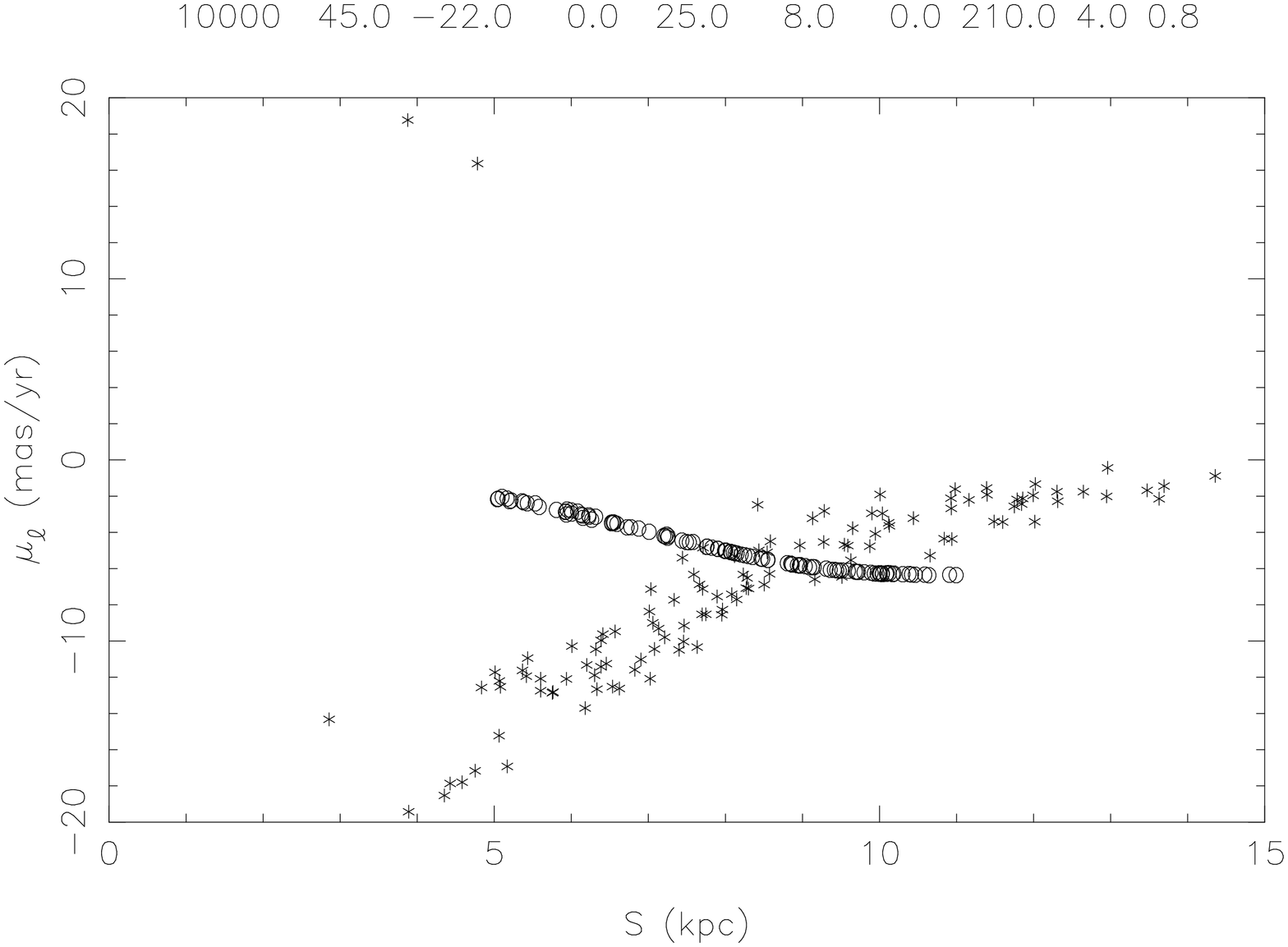,angle=270,height=6truecm}}
\vskip .1truecm
{\psfig{figure=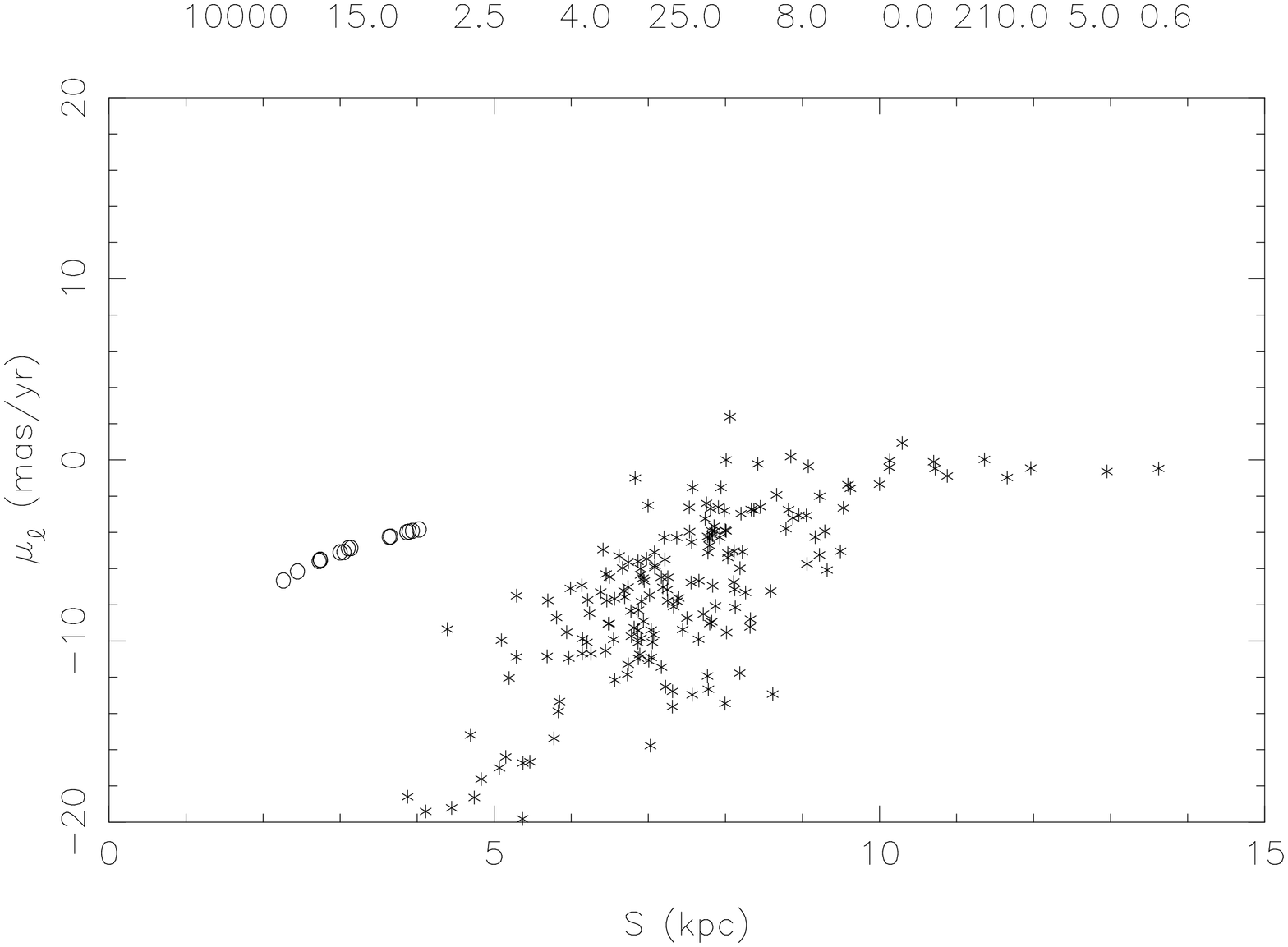,angle=270,height=6truecm}}
\vskip -6truecm
\hskip 7.5truecm
{\psfig{figure=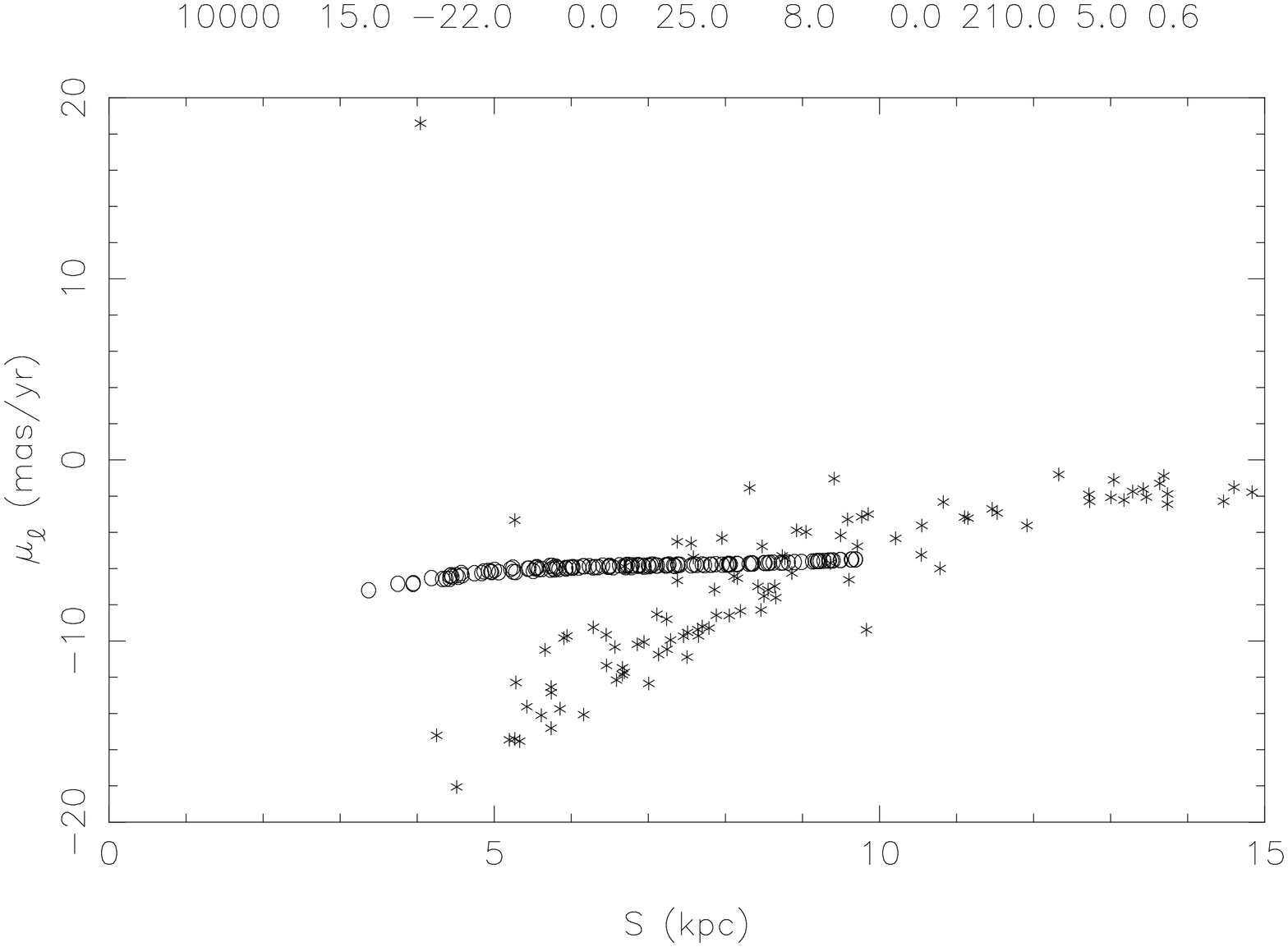,angle=270,height=6truecm}}
\figcaption{
The `observed' longitudinal proper motion as a function of distance
from the Sun is shown for several models.
The top panels are for R75 (left (a)) and R45 (right (b)), the bottom ones for 
E15 (c,d).
All particles are within 0.1 kpc from 
the \losn\ toward Baade's window ($\ell,b$)=(\decdeg2.5,4\degr)
in the left panels and
the negative tangent point ($\ell,b$)=($-$22\degr ,0\degr ) in
the right panels.
Asterisks denote particles from the disk+bar N--body simulation (Fux 1997),
circles the particles drawn from the respective analytic ring densities
(see \Sct 5.1, $\phi_{\rm bar}\equiv \phi_{\rm ring}$).
No velocity disperion is attributed to the ring particles; this
exaggerates the effect seen in the right--hand panels slightly, but
note that a velocity component of 10 \kms\
at a distance of 5 kpc only amounts to a proper motion of 0.4 mas/yr.
The total number of ring particles is 10\% of the number
of N--body particles. 
The velocity of the local standard of rest is taken to be 210 \kms .
}
\end{figure*}

In the right--hand panels of \Fg\DMLR\ we 
show the longitudinal proper motion as a 
function of distance for particles along the \losn\ toward
the negative tangent point ($\ell$=$-$22\degr) for two ring models.
The ring creates a larger distance spread for stars with
similar proper motions, hence a higher \mlod\ for events
with long durations, either via ring self--lensing
or ring--bulge lensing. The effect is somewhat exaggerated
in our simplistic approach without ring velocity dispersion, but
the trend is unambiguous; an added 
velocity dispersion of 50 \kms\ perpendicular to
the \losn\ gives a scatter in proper motions of only 2 mas/yr 
at a distance of 5 kpc. In \Fg\DMLR , we can see that this
would not change the configuration significantly.

Looking at the actual EDD for this \losn\ (\Fg\FTE\&\CFTE a), we see  
that the events are on average longer than toward Baade's window, even
without a ring. Any observed tangent--point EDD would have to
be compared to an EDD along a ``clean'' disk--only \losn , not to 
the EDD measured toward Baade's window. So, compared to the
``no ring'' EDDs in \Fg\FTE a, both ring models show significant
excesses of long events.
In fact, R45 yields longer events than E15, but for a 
narrower range of durations. Self--lensing gives rise to an excess
of 60--100 day events for E15, whereas the extra 90--110 day events in
R45 seem to come from ring--bulge events (\Fg\DMLR ).
When giving the ring particles an isotropic 
velocity dispersion of 50 \kms , the median timeduration does
not change for any of the models considered. 
The 95--\% timeduration changes 
marginally; for the ``Exx'' models it increases by 
about 1 sigma ($\sim$3 days), for the ``Rxx'' models it stays
the same to within 0.5 sigma.

\subsubsection{Toward higher longitudes}

Along a \losn\ beyond the tangent point (\Fg\FTE\&\CFTE b), 
the ring in E15 has no effect on the EDD anymore, while R45 
still gives rise to longer events ($>$50 days) than the others.
In this direction, the EROS team (see Perdereau 1999)
detected two events, of 72 and 98 days, respectively.

Fields at high positive
longitudes were monitored in the MACHO experiment.
Five events were observed at (\decdeg18.77, $-$2\degr ),
with durations of 38, 88, 111, 210 and 242 days,
respectively (MACHO alert webpage, field 301).
In \Fg\FTE\&\CFTE c we see that for $\phi$=15\degr , both with
and without ring, there is an excess of short events in this
direction. For both models with ring, E15 and R45, there is 
a long--event tail and R45 has a significant excess of events
between 40--60 days.

\subsubsection{Toward Baade's window}

Along the \losn\ to the galactic Centre, 
a spiral arm will lower the average transverse velocities 
slightly, via the same deflection as shown in \Fg \STRM . 
With the bulk of the background sources at low transverse
velocities, this would give rise to longer events, possibly 
one step closer towards accounting for the 
high--event--duration tail that is observed and not fully
explained by disk--bulge models (Alcock \etal1997).
However, spiral--arm events do not have large \mlod\ (see \Sct 4).

Ring particles do contribute significantly to the 
\mlod . They can have transverse velocities either higher
or lower than those of the underlying disk, depending
primarily on viewing angle.
For model E15, the transverse velocity of ring particles
at $\ell$=0\degr\ is 
133 \kms , much lower than the local circular velocity ($\sim$170 \kms ).
Hence, E15 should give ring--bulge events that are somewhat
longer than disk--bulge events.
Indeed, \Fg\BWF\ shows that E15 gives 
more events longer than $\sim$50 days than N15.
For higher viewing angles, the transverse velocity at $\ell=0^{\circ}$
of the ring can be very large and give rise to short
events, as for R75 (\Fg\BWF ).
This can also be seen in the left panels of \Fg\DMLR , that show
the transverse proper motions
as a function of distance toward Baade's window, for E15 and R75. 
For E15, the ring particles have the 
same transverse proper motions
as the background distribution.

The EDD for R45 gives a similar long--event tail ($>$65 days) 
to model N15. Apparently, 
the addition of a ring to a bar+disk model has an effect similar
to lowering the viewing angle, not only on the
total \mlod\ but also on the long--event tail of the EDD.
Indeed, for a viewing angle of 45\degr , the elongation does
not have much influence on the proper motions of the ring
particles along Baade's window line of sight and the EDD, 
whereas for 15\degr\ the rounder ring (R15) gives rise to short
events like R75.

Again, with an isotropic velocity dispersion of 50 \kms\ added
to the ring--particle velocities, the median time durations
remain the same, within the error of 0.2 day.
The 95--\% time durations remain the same for the ``Rxx'' models
(with positive observed ring proper motions), with slightly
increased scatter (1.5 days). For the ``Exx'' models, they decrease
somewhat, with decreasing scatter. For eg.~E15, the 
zero--disperion 95--\% timeduration is 91$\pm$3.7 and the
50--\kms\ 95--\% timeduration is 90$\pm$3.4 .
In short, the streaming velocity is by far
the dominant factor in determining time durations, since
the dispersion in the proper motions 
of the foreground particles (ring) is negligible
compared to that of the background particles (bulge).

\section{Conclusions}

We present the first stellar--kinematical evidence
that material in the 3--kpc arm follows closed orbits.
This provides another good reason
to assume that there is an inner ring in the Galaxy, such as
commonly observed in other barred galaxies.
As this ring crosses the \losn\ towards
Baade's window between the observer and the
bulge region, it would be an ideal candidate to explain the high
observed values for the \mlod\ $\tau$ . 
Based on (infra--) red photometry of rings in external barred galaxies, 
we find that a ring could increase the \mlod\ to give values
for $\tau_{-1}$ within 1.5$\sigma$, and for $\tau_{0}$ 
very close to, (recent) observations.
The exact contribution depends on the ring's semi--major axis and viewing
angle and the scaleheight of the disk.

We calculate event--duration distributions for simple models
with a range of viewing angles and along several lines of sight.
For low viewing angles ($<$45\degr ), a ringed model could yield
more long events toward Baade's window than a bar+disk model,
and vice versa.

We show that if the event durations toward the tangent point
prove to be significantly longer than those of general
disk--disk events, this is conclusive evidence in favour
of the ring-- , as opposed to density--wave-- , hypothesis. 
A properly sampled EDD toward the negative--longitude
tangent point, compared to a disk--only EDD, would 
moreover be essential to determine the
values of the ring--density parameters, such as viewing angle and elongation.
Since those parameters are closely connected to the parameters of
the barred potential, this would 
provide an independent way to determine these.

Intermediate--viewing--angle ringed models show equal promise 
for explaining the high values of observed \mlod\ and event--duration
distribution to, and are more realistic than,
low--viewing--angle bar+disk--only models.

\begin{acknowledgements}
We thank Tim Axelrod for helpful discussions and Roger Fux for
providing us with the full
particle positions of his N--body simulation m08t3200.
\end{acknowledgements}

\end{document}